\documentclass[]{article}
\usepackage{amsmath}
\usepackage{natbib}
\usepackage{multirow}
\usepackage{float}

\renewcommand\footnotemark{}

\begin{document}
	
\title{Naive linkage error corrected dual system estimation}

\author{Viktor Ra\v{c}inskij$^{1, 2}$ 
	\footnote{$^{1}$Office for National Statistics, Titchfield PO15 5RR, U.K. (All views expressed are those of the author and do not necessarily reflect the views of Office for National Statistics), $^{2}$Department of Social Statistics and Demography, University of Southampton, Southampton SO17 1BJ, U.K.
	Email: vr1v14@soton.ac.uk. This paper is a part of a PhD project funded by the Office for National Statistics.}}

\maketitle

\section{Introduction}
Capture-recapture is a family of statistical methods that allow estimation of the size of a population from multiple incomplete data sources. These methods are often used in the absence of a sampling frame or when multiple data sources do not cover all members of the population. The use of capture-recapture ranges from estimation of animal populations to census undercoverage \citep{seber82, mccreamor15, wolter86, brownetal19}

The utility of capture-recapture methods is offset by strong underlying assumptions and much of the related research is focused on mitigating failures of those assumptions. In this paper we discuss a failure in the perfect linkage assumption in a special case of capture-recapture known as the dual system estimation. There are already some dual system estimators that allow correction for linkage error \citep{dingfienb94, dictot15, dewolfetal19}. In this paper we propose a naive linkage error corrected dual system estimator. The advantage of the proposed approach is that it permits an easy way to obtain the corresponding variance estimator. The latter may be useful not only for assessing the variance of the linkage error corrected dual system estimator but also when determining a sample size of a rematch study.

\section{Dual system estimator}
The dual system estimator is a special case of capture-recapture estimator that uses information from two samples (also referred to as lists or sources) only. This estimator is well-established and its properties are well-known \citep{wolter86}. We provide just a short recap on the method and related notation.

Let $N$ be an unknown population size that needs to be estimated. Suppose two lists are available / two samples are drawn from the population, with $n_{1+}$ and $n_{+1}$ observed cases in the samples one and two, respectively. The expected values are $E\left(n_{1+}\right) = Np_{1+}$ and $E\left(n_{+1}\right) = Np_{+1}$, where $p_{1+}$ and $p_{+1}$ are the probabilities of being captured in the sample one and two, respectively 

The assumptions are: (a) \textit{closed population} meaning that $N$ does not change between two sampling occasions; (b) \textit{causal independence} meaning that the joint probability of two events equals to the product of marginal list's probabilities of each event; (c) \textit{homogeneity} of inclusion probabilities meaning that the inclusion probabilities in two list are constant or uncorrelated (which means that it is sufficient that inclusion probabilities are constant on one of the lists); (d) \textit{absence of spurious events}, that is only the members of a target population are captured by lists and there are no duplication in captures; (e) \textit{perfect linkage} meaning that the cell count $n_{11}$ of those who are in both sources could be established exactly.

From (b) it follows that $E\left(n_{11}\right) = Np_{1+}p_{+1} = Np_{11}$. Under the above assumptions, the maximum likelihood estimator of $N$ is
\[\hat{N} = \left\lfloor{\frac{n_{1+}n_{+1}}{n_{11}}}\right\rfloor \approx \frac{n_{1+}n_{+1}}{n_{11}},\]
where $\lfloor\cdot\rfloor$ is the greatest integer function \citep{pollock76, wolter86}.

\section{Linkage error corrected dual system estimation}
There have been some development of the estimation techniques that allow to correct or adjust for the linkage error in the dual system estimation. The majority of research in this area follows the idea proposed by \citet{dingfienb94}: to estimate the linkage error rates using a high-quality rematch study on a subsample of the cases used in the dual system estimation and then feed those estimates back into the capture-recapture estimation.

There is a number of simplifying assumptions required by the method of \citet{dingfienb94} and we refer a reader to the original paper. In a nutshell, let the parametrisation be as in the preceding section. In addition, let $\alpha$ be the probability that a correct link is established among the records that truly match, and let $\beta$ be the probability that the incorrect link is declared among the records that truly do not match. Then the probability that a true link is established is $\alpha p_{1+}p_{+1}$ and the probability that an incorrect link is made is $\beta p_{1+}(1 - p_{+1})$. It leads to $p_{11} ^{*} = \alpha p_{1+}p_{+1} + \beta p_{1+}(1 - p_{+1})$ and the remaining cell probabilities can be easily worked out. \citet{dingfienb94} demonstrate that if $\alpha$ and $\beta$ are available, the maximum likelihood estimator of $N$ is
\[\hat{N} ^ {(DF)} = \frac{n}{\hat{p}_{1+} + \hat{p}_{+1} - (\alpha - \beta)\hat{p}_{1+}\hat{p}_{+1} - \beta\hat{p}_{1+}},\] 
where $n = n_{11} + n_{10} + n_{01}$, the number of records in the source 1 or the source 2.

In papers by \citet{dictot15} and \citet{dewolfetal19}  the above estimator is further developed and some of the assumptions are relaxed.

\section{Naive linkage error corrected dual system estimation}
We now introduce an alternative linkage error corrected dual system estimator to the ones considered above. We refer to this estimator as a naive linkage error corrected estimator because it is not the maximum likelihood estimator and does not posses all the properties of the maximum likelihood estimators. However, it has two advantages. The first one, that it is an easy to derive estimator. The second one, that it is straightforward to work out the corresponding variance estimator. 

In what follows, we assume that all the assumptions of the dual system estimator hold except perfect linkage. In addition, all simplifying assumptions from \citet{dingfienb94} also hold. Note that similarly to all the methods discussed, we are working with records rather than matching pairs in our discussion.

Let $\pi$ be the number of unlinked records (false negatives) among the true matches, and $\eta$ be the number of incorrectly linked records (false positives) among the true non-matches. Then using the parametrisation form the previous sections, $E\left(\pi\right) = (1 - \alpha)p_{1+}p_{+1}N$ and $E\left(\eta\right) = \beta p_{1+}(1 - p_{+1})N$ and the observed match count $n_{11} ^ {*}$ that corresponds to the $p_{11} ^{*} = \alpha p_{1+}p_{+1} + \beta p_{1+}(1 - p_{+1})$ is $n_{11} ^ {*} = n_{11} - \pi + \eta$.

Note that the marginal counts $n_{1+}$ and $n_{+1}$ are unaffected by linkage errors. So if a rematch study is conducted and estimators $\hat{\pi}$, $\hat{\eta}$ for $\pi$ and $\eta$ are available, we can estimate $\hat{n}_{11}$ = $n_{11} ^ {*} + \hat{\pi} - \hat{\eta}$ = $n_{11} ^ {*} + \hat{\nu}$, $\hat{\nu} = \hat{\pi} - \hat{\eta}$. If $\hat{\nu}$ is some unbiased estimator for ${\pi} - {\eta}$, then the naive linkage error corrected estimator is defined as
\begin{equation}
	\tilde{N} = \frac{n_{1+}n_{+1}}{n_{11} ^ {*} + \hat{\nu}} \label{eq:nlcdse1}.
\end{equation}
The above estimator is useful for the point estimation. For the variance estimation, we can write ($\ref{eq:nlcdse1}$) as
\begin{equation}
	\tilde{N} ^ {(\epsilon)} = \frac{n_{1+}n_{+1}}{n_{11} + \epsilon} \label{eq:nlcdse2},
\end{equation}
where $E\left(\epsilon\right) = 0$ and $V\left(\epsilon\right) = \sigma_{\epsilon} ^ 2$.

We can estimate $\hat{\nu}$ using the Horvitz-Thompson estimator under a certain sampling design. For instance, under the simple random sampling without replacement if we sampled $n_r$ records from the source 1 in a rematch study, the estimator for $\nu$ would be
\[\hat{\nu} = \frac{n_{1+}}{n_r}\sum_{k \in s}{y_k},\]
where 
$$
y_k =
\begin{cases}
~~1 \text{\small{ if a sampled record is false negative}} \\
-1 \text{\small{ if a sampled record is false positive}} \\
~~0 \text{\small{ otherwise}}.
\end{cases}
$$

If $N$, $n_{1+}$, $n_{+1}$ are large and $\hat{\nu}$ is unbiased, then from the fact that the expected value of the ratio approximately equals to the ratio of expected values it follows that $\tilde{N}$ is approximately unbiased for $N$.

\section{Variance estimation}
The variance of the standard dual system estimator can be approximated using a Taylor series expansion \citep{wolter86}. Expanding $\hat{N}$ around $Np_{1+}$, 
$Np_{+1}$ and $Np_{1+}p_{+1}$ gives
\begin{equation*}
\begin{aligned}
\hat{N} & = \frac{n_{1+}n_{+1}}{n_{11}} = g(n_{11}, n_{1+}, n_{+1}) \\
		& \approx N + \frac{1}{p_{1+}}(n_{1+}  - Np_{1+}) + \frac{1}{p_{+1}}(n_{+1}  - Np_{+1}) - \frac{1}{p_{1+}p_{+1}}(n_{11}  - Np_{1+}p_{+1}).
\end{aligned}
\end{equation*}

Under the multinomial $V\left(n_{1+} \right) = Np_{1+}(1 - p_{1+})$, $V\left(n_{+1} \right) = Np_{+1}(1 - p_{+1})$, $V\left(n_{11} \right) = Np_{1+}p_{+1}(1 - p_{1+}p_{+1})$, $Cov\left(n_{1+}n_{+1}\right) = 0$, $Cov\left(n_{1+}n_{11}\right) = Np_{1+}p_{+1}(1 - p_{1+}p_{+1}) - Np_{1+} ^ 2 p_{+1}p_{+0}$ and $Cov\left(n_{+1}n_{11}\right) = Np_{1+}p_{+1}(1 - p_{1+}p_{+1}) - Np_{1+}p_{+1}p_{0+}p_{+1}$, which allows us to obtain the following variance approximation
\[V\left(\hat{N}\right)\approx N\frac{p_{0+}p_{+0}}{p_{1+}p_{+1}}.\]

Regarding the variance of the naive estimator, we can do the similar expansion as above. However, this time we will have $V\left(\hat{n}_{11} \right) = Np_{1+}p_{+1}(1 - p_{1+}p_{+1}) + \sigma_{\epsilon} ^ {2}$. With some algebra, we get
\begin{equation}
	V\left(\tilde{N}\right)\approx N\frac{p_{0+}p_{+0}}{p_{1+}p_{+1}} + \frac{\sigma_{\epsilon} ^ {2}}{(p_{1+}p_{+1}) ^ 2} \label{eq:var}.
\end{equation}

The proposed variance estimator $\hat{V}\left(\tilde{N}\right)$ is obtained by replacing the parameters in (\ref{eq:var}) with the corresponding estimates:
\begin{equation}
\hat{V}\left(\tilde{N}\right) = \tilde{N}\frac{\hat{p}_{0+}\hat{p}_{+0}}{\hat{p}_{1+}\hat{p}_{+1}} + \frac{\hat{\sigma}_{\epsilon} ^ {2}}{(\hat{p}_{1+}\hat{p}_{+1}) ^ 2} \label{eq:varest},
\end{equation}
where, say, $\hat{p}_{1+} = n_{1+} / \tilde{N}$ and all the remaining parameter estimates are computed in the similar way.

Note that $\sigma_{\epsilon} ^ {2} = V\left(\hat{\nu}\right)$ and it depends on the estimator of $\nu$. Say, in the example above with the Horvitz-Thompson under the simple random sampling without replacements, the variance expression would be the familiar
\[V\left(\epsilon\right) = \sigma_{\epsilon} ^ {2} = n_{1+} ^ 2 \frac{1 - f}{n_r} S_{y} ^ {2} = n_{1+} ^ 2 \frac{1 - f}{n_r} \frac{1}{n_r - 1} \sum_{k}\left(y_k - \bar{y}\right) ^ 2,\]
where $f$ is the sampling fraction. To estimate $\sigma_{\epsilon} ^ {2}$, $S_{y} ^ {2}$ is replaced by the corresponding sample variance.

\section{Simulation study}
We conduct a very basic simulation study to assess whether the proposed linkage error corrected estimator for $N$ and the corresponding variance estimator perform as expected over a range of simple scenarios. Two data sources with varying coverage probabilities are generated from a population with $N = 1000$ and linkage errors are introduced into the initially perfect matching of the elements. A simple random sample without replacements is drawn from the source 1 to mimic a rematch study which detects all the matching errors. Sampling fractions $f = n_{r} / n_{1+} = \{0.1, 0.2\}$ are explored. The Horvitz-Thompson estimator is used to estimate $\hat{\nu}$ which is then fed into the naive linkage error corrected estimator. Each scenario is run over $10000$ iterations.

\begin{table}[!htbp]
	\caption{Simulation results}
	\begin{center}
		{\scriptsize	
			%\hskip -2.3cm
			\begin{tabular}{ r r r r r r r r r r r r }
				\hline\hline
				&  &  &  &  & \multicolumn{3}{c}{\textsc{erb} \%} & \multicolumn{3}{c}{\textsc{erse} \%} & \multicolumn{1}{c}{\textsc{arse} \%} \\ 			
				$p_{1+}$ & $p_{+1}$ & $\alpha$ & $\beta$ & $f$ &  $\hat{N}$ & $\hat{N}_{e}$ & $\tilde{N}$ & $\hat{N}$ & $\hat{N}_{e}$ & $\tilde{N}$ & $\tilde{N}$ \\ \hline
				0.9 & 0.8 & 0.02 & 0.05 & 0.2 & 0.01 & 1.33 & 0.03 & 0.53 & 0.82 & 1.33 & 1.30 \\
				0.9 & 0.8 & 0.02 & 0.05 & 0.1 &      &      & 0.05 &  	  &      & 1.89 & 1.78 \\
				0.9 & 0.8 & 0.05 & 0.02 & 0.2 &      & 4.50 & 0.04 &      & 1.08 & 1.75 & 1.74 \\
				
				0.9 & 0.8 & 0.05 & 0.02 & 0.1 &      &      & 0.05 &      &      & 2.54 & 2.49 \\
				0.9 & 0.8 & 0.05 & 0.08 & 0.2 &      & 4.05 & 0.06 &      & 1.11 & 1.89 & 1.87  \\
				0.9 & 0.8 & 0.05 & 0.08 & 0.1 &      &      & 0.07 &      &      & 2.79 & 2.71 \\
				
				0.8 & 0.7 & 0.02 & 0.05 & 0.2 & 0.02 & 0.24 & 0.06 & 1.03 & 1.32 & 1.96 & 1.92 \\
				0.8 & 0.7 & 0.02 & 0.05 & 0.1 &      &      & 0.07 &      &      & 2.66 & 2.56 \\
				0.8 & 0.7 & 0.05 & 0.02 & 0.2 &      & 4.50 & 0.02 &      & 1.53 & 2.24 & 2.23 \\
				0.8 & 0.7 & 0.05 & 0.02 & 0.1 &      &      & 0.13 &      &      & 3.18 & 3.04 \\
				
				0.8 & 0.7 & 0.05 & 0.08 & 0.2 &      & 2.21 & 0.04 &      & 1.60 & 2.60 & 2.56 \\
				0.8 & 0.7 & 0.05 & 0.08 & 0.1 &      &      & 0.14 &      &      & 3.70 & 3.61 \\
				\hline
			\end{tabular}
		}
	\end{center}
\end{table}

We compare the performance of three estimators: the naive linkage error corrected estimator ($\tilde{N}$), the standard dual system estimator with perfect linkage ($\hat{N}$) and the dual system estimator when linkage errors are present ($\hat{N}_{e}$). The quality is assessed in terms of relative bias and relative standard error. For each estimator a simulated distribution of estimates is used to compute the empirical relative bias (ERB), and the variance of the distribution of estimates is used to compute the empirical relative standard error (ERB). On each simulation iteration variance estimator (\ref{eq:varest}) is applied to estimate the variance of $\tilde{N}$. The resulting distribution of the variance estimates is used to produce the average relative standard error (ARSE) of the variance estimator of $\tilde{N}$.

We are interested in (a) checking whether $\tilde{N}$ is approximately unbiased, (b) how much additional variability is introduced by adjusting the dual system estimator for linkage error, (c) whether the proposed approximate variance estimator (\ref{eq:varest}) has a reasonable performance.

The results summarised in Table 1 confirms that the naive linkage error corrected estimator is approximately unbiased and that the proposed approximate variance estimator produces estimates that are close to the empirical values. The relative standard error of $\tilde{N}$ is two to three times of the dual system estimator with perfect linkage across scenarios considered. Of course, this simulation study is very simplistic and we must be cautious assuming that in real applications the difference in variances between the estimator with perfect linkage and the linkage error adjusted estimator would be as observed here.  

\section{Conclusions and future work}
In this paper we presented a simple linkage error corrected dual system estimator and the corresponding approximate variance estimator. A small simulation study has shown that both estimators perform as expected in a very basic setting.

Future research will explore whether the proposed approaches can be useful in practice.

\end{document}